# Intrinsic charge transport on the surface of organic semiconductors

V. Podzorov [1*], E. Menard [2], A. Borissov[1], V. Kiryukhin[1], J. A. Rogers[2], and M. E. Gershenson[1]

[1] *Department of Physics and Astronomy, Rutgers University, Piscataway, New Jersey*

[2] *Department of Materials Science and Engineering, University of Illinois, Urbana-Champaign, Illinois*

03.01.2004

The novel technique based on air-gap transistor stamps enabled realization of the intrinsic (not dominated by static disorder) transport of the electric-field-induced charge carriers on the surface of rubrene crystals over a wide temperature range. The signatures of the intrinsic transport are the anisotropy of the carrier mobility, $\mu$, and the growth of $\mu$ with cooling. The anisotropy of $\mu$ vanishes in the activation regime at lower temperatures, where the charge transport becomes dominated by shallow traps. The deep traps, deliberately introduced into the crystal by X-ray radiation, increase the field-effect threshold without affecting the mobility. These traps filled above the field-effect threshold do not scatter the mobile polaronic carriers.

*Electronic mail: *podzorov@physics.rutgers.edu*

One of the central problems in the experiments with organic semiconductors is realization of the *intrinsic, not limited by static disorder* charge transport. The difficulty of its demonstration is related to the polaronic nature of the charge carriers in organic molecular crystals (see, e.g., [1]): small polarons can be trapped by numerous types of crystal defects. The benchmark in the study of the intrinsic transport has been established in the time-of-flight (TOF) experiments, in which the drift velocity of photo-generated charge carriers has been measured in strong electric fields [2]. These experiments address the charge transport in the bulk of ultra-pure organic crystals at small carrier densities, $n$. The intrinsic transport in TOF experiments is characterized by an increase of the carrier mobility with cooling and a pronounced anisotropy of the mobility with respect to the crystal orientation. Theoretical treatment of these results was based on the Holstein concept of small polarons (see, e.g. [1,3]).

Until recently, the experimental tool to probe the intrinsic charge transport on the surface of organic semiconductors was unavailable: the transport in the organic thin-film transistors is typically dominated by disorder [4,5]. Fabrication of the single-crystal organic field-effect transistors (OFETs) (see, e.g., [6,7,8]) provides an opportunity to study the charge transport on the organic surface with significantly reduced disorder. It also offers a possibility to explore the regime of high charge densities, many orders of magnitude greater than in the TOF experiments, which may lead to observation of new electronic phases. However, despite recent technological advances realization of the intrinsic transport on the organic surface remains a challenge, because the concentration of defects at the surface is typically higher than in the bulk.

In this Letter, we report on observation of the intrinsic transport of field-induced charges on the surface of single crystals of rubrene. Application of a novel experimental technique based on the "air-gap" transistor stamps [9] allowed realization of a high room-temperature mobility for *p*-type carriers (~20 cm$^2$/Vs). Two signatures of the intrinsic transport, - the mobility increase with decreasing temperature and the mobility anisotropy, have been observed in the temperature range ~150 – 300 K. At lower temperatures, where the charge transport is dominated by shallow traps, the mobility decreases exponentially with cooling and anisotropy of $\mu$ vanishes. Our experiments elucidate the role of shallow and deep traps in the field-effect experiments with organic semiconductors.

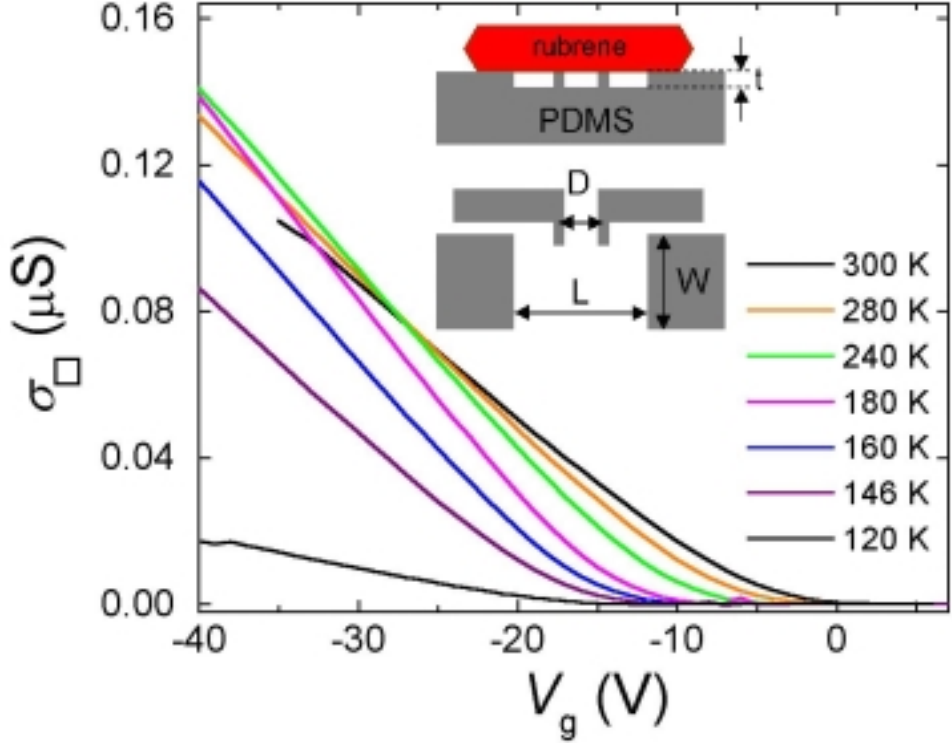

**Fig. 1.** The surface conductivity per square, $\sigma_\square$, as a function of the gate voltage, $V_g$, measured along the **b** crystallographic direction of a rubrene single crystal at different temperatures with the 4-probe "air-gap" transistor stamp (the source-drain voltage $V_S = 5$ V). The inset schematically shows the transistor stamp: $L = 0.75$ mm, $D = 0.25$ mm, $W = 1.25$ mm, $t = 5$ μm; $C_i = 0.2$ nF/cm$^2$.

High-quality single crystals of rubrene have been grown from the vapor phase in a stream of ultra-pure hydrogen in a horizontal reactor [6,10]. We have used a novel design of transistor stamps, based on flexible elastomeric substrate (polydimethylsiloxane, or PDMS) with a surface relief pattern, on which the source, drain and gate electrodes are deposited in a single vacuum cycle by flash deposition of 2 nm of Ti (for better adhesion) and 20 nm of Au (inset in Fig. 1) [9]. The gate electrode is recessed from the level of source/drain contacts by 3 to 5 μm. The continuous metal film breaks at the vertical walls of the pattern imprinted on the PDMS surface: this provides electrical isolation between the electrodes. Fabrication of the field-effect structures is completed after the organic crystal is laminated against the surface of the stamp [11]. Since the gate electrode is recessed from the level of the contacts, the stamp does not touch the crystal surface within the area of the conduction channel; this allows us to minimize the density of surface defects introduced in the process of FET fabrication. The electrical breakdown strength of these structures is very high ($\geq 1.5 \times 10^5$ V/cm) even at the atmospheric pressure, in line with the prediction of Paschen's law for a micron-size separation between the gate electrode and the contacts (see, e.g. [12]).

In OFETs, the charge carriers are injected into the semiconductor through the Schottky barriers at the metal/organic interface [13]. In order to exclude the contact effects, we have used the 4-probe contact configuration [6]. Figure 1 shows the dependence of the sheet conductance $\sigma_\Box = (D/W)(I_{SD}/V)$ on the gate voltage $V_g$ measured at a constant source-drain voltage $V_S$ ($I_{SD}$ is the source-drain current, $V$ is the voltage difference between the voltage probes separated by a distance $D$, and $W$ is the channel width; the drain is grounded). At large $V_g$, the dependences $\sigma_\Box(V_g)$ are linear: this regime corresponds to the $V_g$-independent mobility of carriers $\mu = \sigma_\Box/en$ ($n$ is the two-dimensional density of mobile field-induced carriers). The carrier mobility in the linear regime is proportional to the slope of $\sigma_\Box(V_g)$ dependences, $\mu = \left(\frac{1}{C_i}\right)\left(\frac{d\sigma}{dV_g}\right)$ [14], where $C_i$ is the specific capacitance between the gate electrode and the channel (the gate dielectric is air or vacuum, depending on the experimental conditions, with dielectric constant $\varepsilon=1$). For the estimate of $\mu$, we have used the capacitance calculated from the device geometry; this value is consistent with the direct measurements of $C_i$ in the test structures formed by lamination of the stamp against a metallic surface [9]. This definition of $\mu$ assumes that all charge carriers with the density $n = \frac{C_i \cdot (V_g - V_g^{th})}{e}$, induced by the transverse electric field above the threshold, are mobile. A very weak dependence $\mu(V_g)$ observed for our field-effect structures, justifies this assumption. For comparison, the mobility in amorphous silicon ($\alpha$-$Si:H$) FETs [15] and organic TFTs [16] is strongly $V_g$-dependent. In these cases, most of the charge above the field-effect threshold is induced into the in-gap localized states, and the conduction is governed by the multiple trap and release (MTR) mechanism.

Figure 1 shows the evolution of the $\sigma_\Box(V_g)$ dependences measured along the ***b*** crystallographic direction of rubrene with decreasing temperature. The mobility, proportional to the slope $d\sigma_\Box/dV_g$ initially increases, reaches a maximum at $T \sim 150$ K, and decreases rapidly with further cooling (see also Fig. 2). The mobility measured along the ***a*** direction is found to be systematically lower by a factor of 2.5 - 3 at 300 K, in line with our recent study [8]; this anisotropy is due to a stronger π-π overlap along the *b*-direction in rubrene crystals. The mobility anisotropy persists with cooling down to $T \sim 150$ K. The field-effect threshold increases with decreasing temperature (Figs. 1 and 2) and precludes measurements of $\mu$ below 100 K [17].

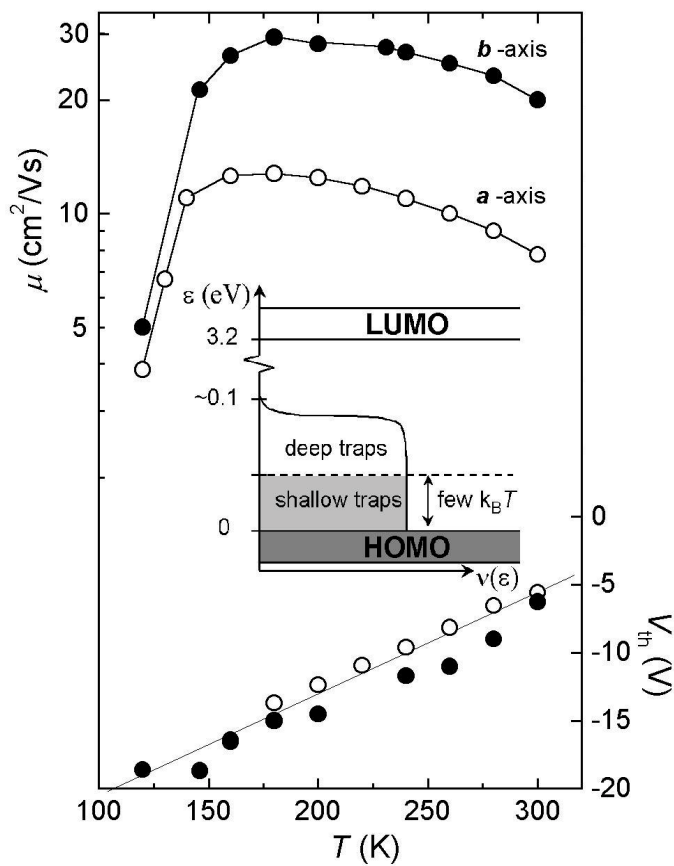

**Fig. 2.** The temperature dependence of the field-effect mobility and the threshold voltage measured along the *a* and *b* crystallographic directions of rubrene using the 4-probe "air-gap" transistor stamp. The inset schematically shows the electronic states near the HOMO energy in the studied rubrene crystals.

On the basis of our experimental results, we can qualitatively reconstruct the energy diagram for the electronic states near the HOMO level in the studied rubrene crystals (see the inset to Fig. 2). The HOMO and LUMO levels in rubrene are separated by $\sim 3.2$ eV energy gap. Within the gap, there are localized electronic states (traps) associated with the crystal defects, such as chemical impurities, structural disorder and surface states. Injection of *p*-type carriers results in filling the traps and shifting the Fermi energy at the organic surface, $E_F$, towards the HOMO level. Below the field-effect threshold ($|V_g| < |V_{th}|$), the injected charge is trapped in the localized states with energies separated by more than a few $k_BT$ from the HOMO level (the *deep* traps). When Fermi level reaches the traps with energies within a few $k_BT$ from the HOMO level (the *shallow* traps), the surface conductivity increases dramatically, by many orders of magnitude, owing to the thermal excitation of the carriers from the shallow traps to the HOMO level.

A small field-effect threshold at 300K (see Fig. 2) indicates that the concentration of deep traps with energies $\geq 0.1$ eV above the HOMO level, $N_{tr} = \dfrac{C_i \cdot V_{th}}{e}$, does not exceed $0.7 \times 10^{10}$ cm$^{-2}$. With cooling, the borderline between deep and shallow traps shifts towards the HOMO level. The increase of $V_{th}$ with decreasing temperature suggests that the total density of deep traps increases to $N_{tr} = 2 \times 10^{10}$ cm$^{-2}$ as $T$ decreases from 300 to 150 K [18]. Interestingly, the quasi-linear increase of $V_{th}$ with cooling suggests that the energy distribution of trap states, $\partial N_{tr}/\partial E = (C_i/k_B e)(\partial V_{th}/\partial T) \approx 10^{12}$ cm$^{-2}$eV$^{-1}$, is almost energy-independent within ~0.1 eV near the HOMO level. This observation is in a sharp contrast with an exponential energy distribution of the in-gap states in the α-*Si:H* FETs [15]. In the latter case, the density of shallow traps is so high that $E_F$ remains within the tail states even at the highest accessible gate voltages, and the carrier mobility depends strongly on $V_g$ and decreases exponentially with decreasing $T$.

The $\mu(T)$ dependences measured along the *a* and *b* axes are shown in Fig. 2. The room-temperature value of $\mu^b \sim 20$ cm$^2$/Vs along the *b* axis in our rubrene devices exceeds by a factor of 10 $\mu(300K)$ observed in the TOF experiments with naphthalene, anthracene, and perylene [2]. Note that the electric field along the conduction channel in our experiment (~ 50 V/cm) is by several orders of magnitude smaller than the typical fields in the TOF measurements. Therefore, de-trapping due to the large drag-fields can be neglected.

Two transport regimes are clearly seen on Fig. 2: (a) at high temperatures, $T = 150 - 300$ K, the mobility is strongly anisotropic and increases with cooling, (b) at $T < 150$ K, the mobility rapidly decreases with cooling, and the anisotropy of mobility vanishes. The low-temperature drop of the mobility can be fitted (over a limited range 100 - 150 K) by an activation dependence $\mu(T) = \mu_0 \exp(-T_0/T)$ with the activation energy $k_B T_0 \sim 70$ meV, which is the same for both crystallographic directions (see also the inset to Fig. 4). We argue that the former regime corresponds to the intrinsic transport of polaronic charge carriers, whereas at low temperatures the charge transport is dominated by the multiple trapping and release of carriers by shallow traps [16]. The increase of $\mu$ with cooling has been typically observed for devices with $\mu^b$(300K) $\geq$ 10 cm$^2$/Vs. For devices with $\mu^b$(300K) $\leq$ 10 cm$^2$/Vs, $\mu$ is almost $T$-independent at high $T$, and exponentially decreases at lower $T$ with $T_0$ similar to that for the higher-$\mu$ devices (Fig. 4). These differences might be associated with variations of the surface morphology of the crystals (e.g., the density of steps).

Observation of the signatures of the intrinsic transport at high $T$ does not imply that the trapping processes are completely eliminated. On the contrary, the higher the temperature, the higher the total number of shallow traps involved in the trap-and-release processes. However, at high $T$, the time that a polaron spends within a shallow trap with energy $E_{tr}$, $\tau_{tr} \propto \exp(E_{tr}/k_BT)$, might be much smaller than the time it propagates between the traps, $\tau$. If this is the case ($\tau_{tr} \ll \tau$), the effective drift mobility in the MTR model [19], $\mu_{eff} = \mu_0 \left( \dfrac{\tau}{\tau + \tau_{tr}} \right)$, reduces to the intrinsic (trap-free) mobility $\mu_0$.

In the opposite limit ($\tau_{tr} \gg \tau$), the charge transport is dominated by trapping and $\mu_{eff} = \mu_0(\tau/\tau_{tr}) \propto \exp(-E_{tr}/k_BT)$. This regime is observed for the studied OFETs at $T < 150$ K. The exponential drop of $\mu$ with decreasing $T$ in this regime is governed mainly by the exponential increase of $\tau_{tr}$. The activation energy in the Arrhenius-like dependence, $k_BT_0$ (see inset to Fig. 4), is the integral characteristic of a broad distribution of shallow traps rather then a single trap level. Note that the crossover from the intrinsic to the thermally-activated transport in the bulk has been observed in the TOF measurements of organic crystals with *ppm* impurity concentrations that is close to the trap density in the studied rubrene crystals [20].

Our experiments show that the mobility anisotropy vanishes in the trap-dominated transport regime (see Fig. 2). This observation is also consistent with the MTR model. Indeed, the time of propagation of the carrier between the traps along a certain crystallographic direction, $\tau$, is inversely proportional to the intrinsic mobility in this direction (e.g., $\tau^b \propto 1/\mu_0^b$). Thus, with isotropic distribution of traps in the crystal, one might expect vanishing of the anisotropy of $\mu_{eff}$ in the regime $\tau_{tr} \gg \tau$.

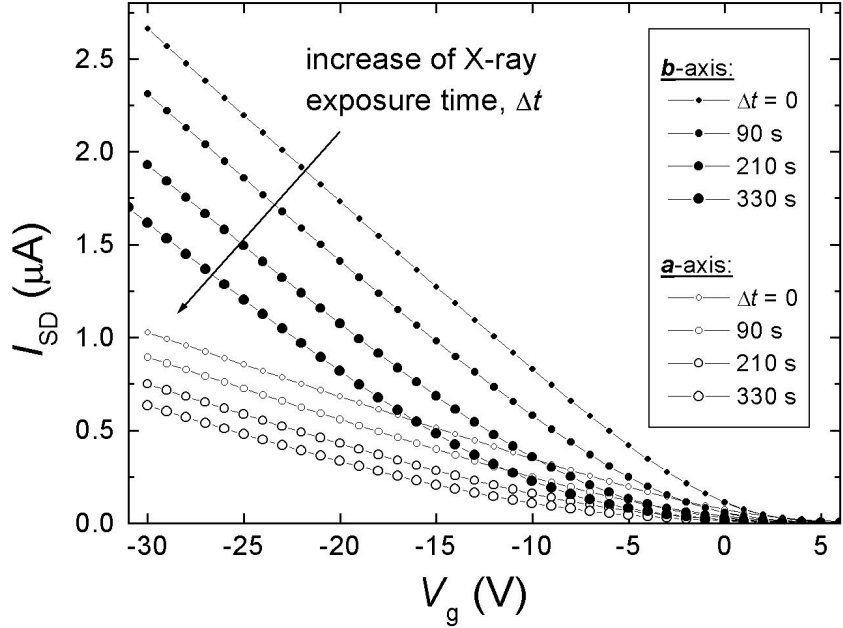

**Fig. 3.** The effect of the X-ray radiation on the trans-conductance characteristics of rubrene OFETs. The $I_{SD}(V_g)$ dependences have been measured simultaneously along the ***a*** and ***b*** crystallographic directions at 300 K using a 2-probe air-gap transistor stamp with two channels oriented perpendicular to each other. After each X-ray exposure, the crystal was measured in air and in the dark; $\Delta t$ indicates the cumulative exposure time. The mobility, proportional to the slope of $I_{SD}(V_g)$, is not significantly affected by X-ray, whereas the field-effect threshold noticeably increases. The initial device characteristics do not recover with time.

In order to understand better the charge trapping in organic semiconductors, we deliberately introduced defects in the rubrene crystals by exposing them to the X-ray radiation [21]. The ionizing radiation breaks the rubrene molecules and produces new chemical species which act as local defects in the crystal structure [22]. The trans-conductance characteristics of the rubrene single-crystal OFET measured along the ***a*** and ***b*** directions before and after the X-ray exposure are shown in Fig. 3; the corresponding dependences $\mu(T)$ and $V_{th}(T)$, measured by the 4-probe technique, are shown in Fig. 4. It is clear that the X-ray treatment significantly increases $V_{th}$, and, thus, the density of deep traps, without affecting $\mu$. The temperature dependence of the mobility is also unaffected. This indicates that (a) the X-ray does not generate shallow traps (within ~ 0.1 eV from the HOMO level), and (b) the deep traps, being filled above the threshold, do not affect the properties of mobile polaronic carriers.

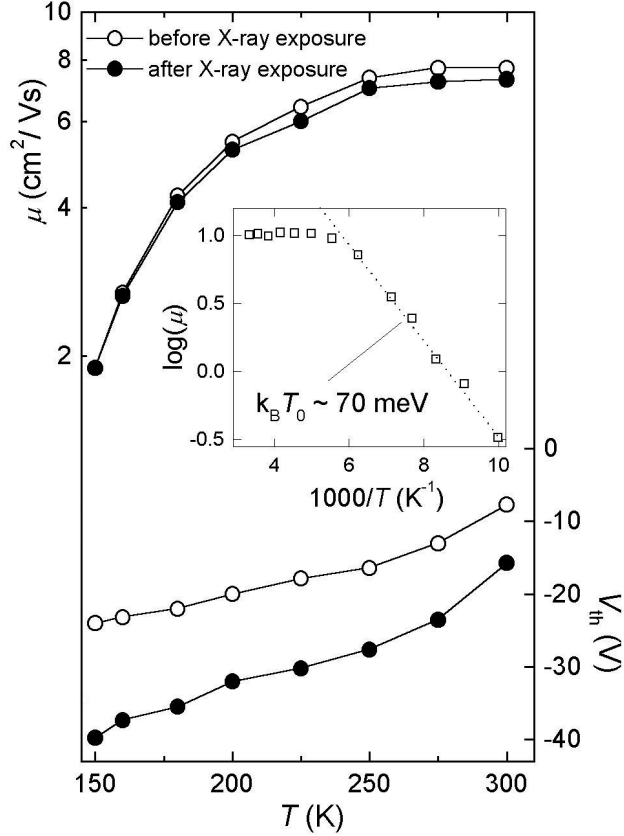

**Fig. 4.** The temperature dependence of the field-effect mobility $\mu(T)$ and the threshold voltage $V_{th}(T)$ of the rubrene OFETs measured by the 4-probe vacuum-gap transistor stamp along the ***b*** axis before (open circles) and after (filled circles) the sample was exposed to X-ray radiation for $\Delta t = 4$ min. The inset is an Arrhenius plot of the mobility along the ***b*** axis for another device (not exposed to the X-rays): the low-*T* data can be fitted by the exponential $\mu(T)$ dependence with the activation energy ~ 70 meV.

To summarize, we have studied the transport of *p*-type carriers on the surface of organic single crystals of rubrene by means of the air-gap field-effect transistor stamps. This technique allowed to minimize the density of surface defects/traps and to realize the intrinsic polaronic transport on the surface of organic semiconductors. Two transport regimes have been identified: (a) the intrinsic regime observed at high temperatures, where the mobility is anisotropic and increases with decreasing temperature, and (b) the shallow trap dominated regime, where the mobility decreases rapidly with cooling, and anisotropy of the mobility vanishes. Our experiments with X-ray treatment of the organic crystals show that the deep traps filled with carriers do not contribute to the scattering of the mobile polarons.

This work at Rutgers was supported by the NSF grant DMR-0405208.